\documentclass[useAMS,usenatbib]{mn2e}

\def\src{SGR\,0418$+$5729}

\def\nh {$N_{H}$}

\def\ergscm2 {erg\,s$^{-1}$cm$^{-2}$}
\def\ss {s\,s$^{-1}$}
\def\cm2 {cm$^{-2}$}

\usepackage{graphicx}
\usepackage{url}
\usepackage{color}
\usepackage{enumerate}

\title[\src ~ as an evolved Quark-Nova compact remnant]{\src ~ as an evolved Quark-Nova compact remnant}
\author[Ouyed et al. (2011)]{R. Ouyed\thanks{E-mail:
rouyed@ucalgary.ca (RO)}, D. Leahy and B. Niebergal \\
Department of Physics and Astronomy, University of Calgary,
2500 University Drive NW, Calgary, Alberta, T2N 1N4 Canada}
\begin{document}

\date{Accepted ---. Received ---; in original form ---}

\pagerange{\pageref{firstpage}--\pageref{lastpage}} \pubyear{2010}

\maketitle

\label{firstpage}

\begin{abstract}
 Soft gamma repeaters and anomalous X-ray pulsars are believed to be magnetars, i.e. neutron
stars powered by extreme magnetic fields, $B\sim10^{14}$-$10^{15}$
Gauss.   The recent discovery of  a soft gamma repeater with low magnetic field ($< 7.5\times 10^{12}$\,Gauss), \src , which shows bursts similar to those of
 SGRs,  implies that a high surface dipolar
magnetic field  might not be necessary for magnetar-like
activity.  We show that the quiescent and bursting properties of
 \src~ find natural explanations in the context of  low-magnetic field Quark-Nova (detonative
  transition from a neutron star to a quark star) remnants, 
  i.e. an old quark star surrounded by degenerate (iron-rich) Keplerian ring/debris ejected during the
   Quark-Nova explosion.  We find that a 16 Myr old quark star surrounded by
    a $\sim 10^{-10}M_{\odot}$ ring, extending in radius  from $\sim 30$ km to $60$ km,
     reproduces many observed properties of \src.
   The SGR-like burst is caused by magnetic
    penetration of the inner part of the ring and subsequent accretion.
     Radiation feedback  results in months-long accretion from
      the ring's non-degenerate atmosphere which matches well
       the observed  decay phase. 
       We make specific predictions (such as an
        accretion glitch of $\Delta P/P \sim - 2\times 10^{-11}$ during
         burst and  a sub-keV proton cyclotron line  from the ring)
        that can be tested by sensitive observations.
\end{abstract}


\section{Introduction}

Soft $\gamma$-ray Repeaters (SGRs) are sources of recurrent, 
short ($t \sim 0.1\,\mathrm{s}$), intense ($L \sim 10^{37-42}~\rm{ergs}$) 
bursts of $\gamma$-ray emission with an energy spectrum characterized by temperatures of $\sim 100$  keV.
 Occasionally SGRs enter into active episodes producing many short X-ray bursts; 
extremely rarely (about once per 50 years per source), SGRs emit a giant flare, 
an event with total energy at least 1000 times higher than their typical bursts. 
The normal pattern of SGRs is intense activity periods 
which can last weeks or months, separated by quiescent phases 
lasting years or decades. 
Current theory explains this energy release as the result of a catastrophic 
reconfiguration of a magnetar's magnetic field. 
AXPs are similar in nature but with a somewhat weaker intensity and
no recurrent bursting.
Several SGRs/AXPs have been found to be X-ray pulsars with 
unusually high spin-down rates, 
usually attributed to magnetic braking caused by their super-strong 
magnetic field.  In all sources with magnetar-like activity, the dipolar fields
span $5\times10^{13}\,{\rm G} <B< 2\times10^{15}\,{\rm G}$, which
is $\sim 10$--1000 times the average value in radio pulsars. Magnetar-like activity previously 
was observed only
in sources with dipolar magnetic fields stronger than
 the electron quantum field, $B_{\rm
Q}=m_e^2c^3/e\hbar\sim 4.4\times10^{13}$ Gauss.

\src\, was discovered on 5 June 2009 when
the Fermi Gamma-ray Burst Monitor (GBM) observed two
magnetar-like bursts (\cite{vanderhorst10}). Follow-up
observations with several x-ray satellites show that it has x-ray
pulsations at $\sim$9.1\,s, well within the range of periods of SGR
sources (\cite{gogus09,esposito10}).
The implied  upper limit on the period derivative of \src\ of $\dot{P} <
6.0\times10^{-15}$\ss\  is by far the
smallest of all known SGRs/AXPs. The corresponding limit on the surface dipolar magnetic
field of \src\, is $B < 7.5\times10^{12}$\,Gauss, making it the
SGR with the lowest surface dipolar magnetic field yet. 
 The upper limit on the period derivative implies a characteristic age of the source in excess of 
 $  > 24$\, Myr in the  standard dipole model, $P/(2\dot{P})$. Despite such a low surface magnetic field ($< <B_{\rm Q}$),   \src\ exhibits all the typical characteristics of an SGR. Its bursting properties can be summarized as follows (\cite{vanderhorst10,gogus09,esposito10}):

\begin{itemize}

\item At a distance of 2 kpc, 
  assuming that the source is located in the Perseus arm,
the estimate energies of the two observed bursts (in the 8-200 keV range) are 
  $4\times 10^{37}$ erg and  $ 2\times  10^{37}$ erg, 
which is at the lower end of the distribution
compared to other SGR bursts
but at the high end for AXP ones. 

\item An optically-thin thermal bremsstrahlung  provides the best fit in both bursts (see Table
2 in \cite{vanderhorst10}). The spectrum softens  from $kT\sim 33$ keV in the frist burst to $kT\sim 20$ keV 
 in the second burst in the time period of $\sim 20$ minutes which separated the two bursts.

\item Immediately following the burst and for the first 160 days (before
 it disappeared behind the sun), \src ~ flux  declined by an order of
  magnitude from 
  $\sim 3\times 10^{-11}$ erg cm$^{-2}$ s$^{-1}$ ($L_{\rm X}\sim1.4\times 10^{34}$ erg s$^{-1}$ ) to $\sim 3\times 10^{-12}$ erg cm$^{-2}$ s$^{-1}$ ($L_{\rm X}\sim 1.4\times 10^{33}$ erg s$^{-1}$). The corresponding blackbody
   temperature declined from 1 keV to 0.8 keV (see Figure 2 in \cite{esposito10}). 
   
\item   The burst luminosity is $\sim 10^{39}$ erg s$^{-1}$ with burst temperature
        in the 20-30 keV range.  Assuming blackbody emission the emitting area can
         be estimated to be on average $A_{\rm b}\sim 8.5\times 10^8$ cm$^2$. 

\end{itemize}

\src's~  current properties (i.e.  following the bursting era) show a 
 spectrum which is well
fit by an absorbed blackbody with a line-of-sight absorption
\nh$=(1.5\pm1.0)\times10^{21}$\cm2 \, and $kT=0.67\pm0.11$\,keV. 
 Using the current
          luminosity of $6\times 10^{31}$ erg s$^{-1}$, and a blackbody temperature
           of 0.67 keV (\cite{rea2010}) the emitting area is $A_{\rm q}\sim 3.1\times 10^{8}$ cm$^2$ 
           ($<< 4\pi (10\ {\rm km})^2$)
             indicative of a hot spot on the surface of the star.

Since the internal field strength required to produce
  crustal cracking in the Magnetar model should be typically in excess of $10^{14}$\, Gauss
  (\cite{td95}),  one wonders how  \src,
    with its weak surface magnetic field, can harbor a much stronger magnetic
    field in its crust.   In fact, for such an old source ($> 24$ Myr) it would take an even stronger
     internal field to crack  the cold crust\footnote{However there are no observational constraints on multipole moments of the surface field or
 on  the internal toroidal magnetic field to rule out the crust cracking model.  
  The tearing mode instability model  proposed for magnetic field
decay could account for short times scale bursts ($< 1$ yr) if the dissipation scale is much shorter than 
typical crust scale ($<< 10^4$ cm) (\cite{lyutikov2003}).}. Could such a difference between the
       surface and crustal field be possible and sustainable? What mechanism
        could lead to such a gradient, and if it exists does it mean that some 
  ordinary pulsars are dormant SGRs waiting to erupt?  Maybe a strong magnetic
 field  is not necessary to explain SGR behavior. 
 The existence of radio pulsars with $B > B_{\rm Q}$  and, so far,  showing only normal
behavior (\cite{kaspi10}) is another clue that magnetic fields larger than the quantum electron field alone may not be
a sufficient condition for the onset of magnetar-like activity.

    Here we present an alternative model which offers natural answers to these outstanding questions.
     In our model high magnetic field strength  is not
  necessary to explain the  bursting phase of SGRs and AXPs : It involves an aligned rotator (a quark star; hereafter QS) and iron-rich debris
    material in the close vicinity ($\sim 20$-$100$ km) of  the QS star. The  QS is the compact remnant of 
    the Quark-Nova (QN) explosion, a detonative 
    transition from a neutron star (NS) to a QS (\cite{ouyed2002,vogt2004,niebergal2010b}).  
    The QN detonation also leads to  ejection of the NS 
outer layers (\cite{keranen2005}). 
 If the QS is born slowly rotating, then the debris formed from the QN ejecta will be in 
co-rotation with the star's dipole field (\cite{oln1}, herefafter OLNI). Sources born with faster rotation   will confine the
debris into a Keplerian ring  at 20-100 km away from the star (\cite{oln2}, hereafter OLNII)\footnote{This   ring is
 unlike a fall-back disk  around a neutron star  (e.g. \cite{trumper2010}). 
  The ring is iron-rich, very close to the star and degenerate.  Similar ring formation
          when a neutron star is born 
         appears implausible since the proto-neutron star  is too large. Later on, there is no 
          mechanism to eject degenerate material unless a violent change
          of state, like a QN occurs.}. In our model SGRs are QS with a co-rotating shell
 while AXPs are QS with a Keplerian ring.  The  debris consists 
 of $\sim 10^{-6}M_{\odot}$ of iron-rich degenerate material. The initial QS surface magnetic field is $10^{15}$ G.   Such initial extreme surface magnetic fields are  natural values for
    quark stars experiencing color ferromagnetism  before they
    enter the color superconducting phase (\cite{iwazaki2005}).
    
    The paper is organized as follows.  Section 2 is a summary of our previous works and sets the stage
 for this work. Section 3 is  the application of our model to evolved accreting QS-ring systems. 
 The thermal feedback between the QS and the ring depends critically on the ring geometry, 
 which evolves from  thick to thin.   For old sources,   the resulting behavior is 
 in a different regime than considered in previous papers. This leads to  new forms  of the equations, which we then fit  to
 the observations.  We conclude in Section 4.
          
    \section{SGRs and AXPs in our model}

       \subsection{The aligned rotator and the vortex band}

     The QS  is born an aligned rotator
     since the superconducting QS confines the interior magnetic field  to vortices aligned with the rotation axis\footnote{The aligned magnetic field of the QS  provides a natural explanation for the lack of persistent radio pulsation  (\cite{ouyed2004,ouyed2006}).} (\cite{ouyed2004,ouyed2006}). As the star spins down via EM
      emission, vortices (and their confined magnetic filed) are expelled, leading
       to magnetic field reconnection at the surface of the star.
                The  period  and magnetic
           field during spin-down   evolve as (\cite{niebergal2006})
           $B = B_0 (1+\frac{t}{\tau_{\rm 0}})^{1/6}$ and 
          $P= P_0 \left(1+ \frac{t}{\tau_{\rm 0}}\right)^{1/3}$. Here  
             $\tau_{\rm 0}= 840\ {\rm s}\ P_{\rm 0, ms}^2/B_{\rm 0, 15}^2$  with  birth
              period and magnetic field  in millisecond and $10^{15}$ G, respectively.
              The period derivative is  $\dot{P}= \dot{P}_0 \left(1+ \frac{t}{\tau_{\rm 0}}\right)^{-2/3}$
               with $\dot{P}_0 = P_0/(3\tau_0)$ and 
                characteristic age $t_{\rm age}=P/3\dot{P}$  different from  $t_{\rm age}=P/2\dot{P}$ in the
                 standard dipole model because vortex expulsion  changes the magnetic braking index from $n=3$
                   (the theoretical value for an aligned spinning dipole) to $n=4$.
                   This
 gives $t_{\rm age} > 16$ Myr for \src. After 16 Myr of spin-down, the  initial magnetic field  $\sim 10^{15}$ G
                will be reduced to $\sim 10^{13}$ G.

      Vortex expulsion   leads to an X-ray
       luminosity given by $L_{\rm X, v}\sim 2\times 10^{35}\ {\rm erg\ s}^{-1} \eta_{\rm X}\dot{P}_{-11}^2$
        where the subscript  ``${\rm v}$" stands for vortex and $\eta_{\rm X}$ is an efficiency parameter
         (see OLNI and
          \cite{ouyed2006}); $\dot{P}$ is in units of $10^{-11}$ s s$^{-1}$.  Shown in Figure \ref{fig:bands} is the $L_{\rm X}$-$\dot{P}$ diagram 
           with the  vortex band shown  (\cite{ouyed2006}; OLNI).   
           
             QS-shell  and non-accreting QS-ring systems evolve along the vortex band.  
              QS-ring systems (AXPs)  enter an accretion era
              where their steady X-ray luminosity is dominated by accretion  
              (OLNII).
              These systems  will evolve along the accretion band (see Figure \ref{fig:illustration}), 
                 and  return to the vortex band (a vortex-dominated phase) much later
     once the ring is consumed (\cite{oln3}, hereafter OLNIII). In OLNIII
         RRATs are explained
        as  the result of AXPs descending back to the vortex band.

\subsection{Transient AXPs in our model}
        
      The QS-Ring system for AXPs 
      allows us to differentiate between AXPs and transient AXPs.  
             Transients AXPs are QS-Ring systems which have not yet entered the
   accretion-dominated phase and are in the vortex band (\cite{oln4}, hereafter OLNIV). Transient AXPs    burst only when the inner wall of the ring
       is permeated by the magnetic field and  is accreted.
       Only during bursting phase does the ring  accrete (OLNIV).
       During quiescence, these are  much quieter than SGRs (although in similar region in the vortex band) 
       since  the ring is more stable than the co-rotating shell (OLNI\&OLNII). 
       
\begin{figure*}
\begin{center}
\includegraphics[width=0.85\textwidth]{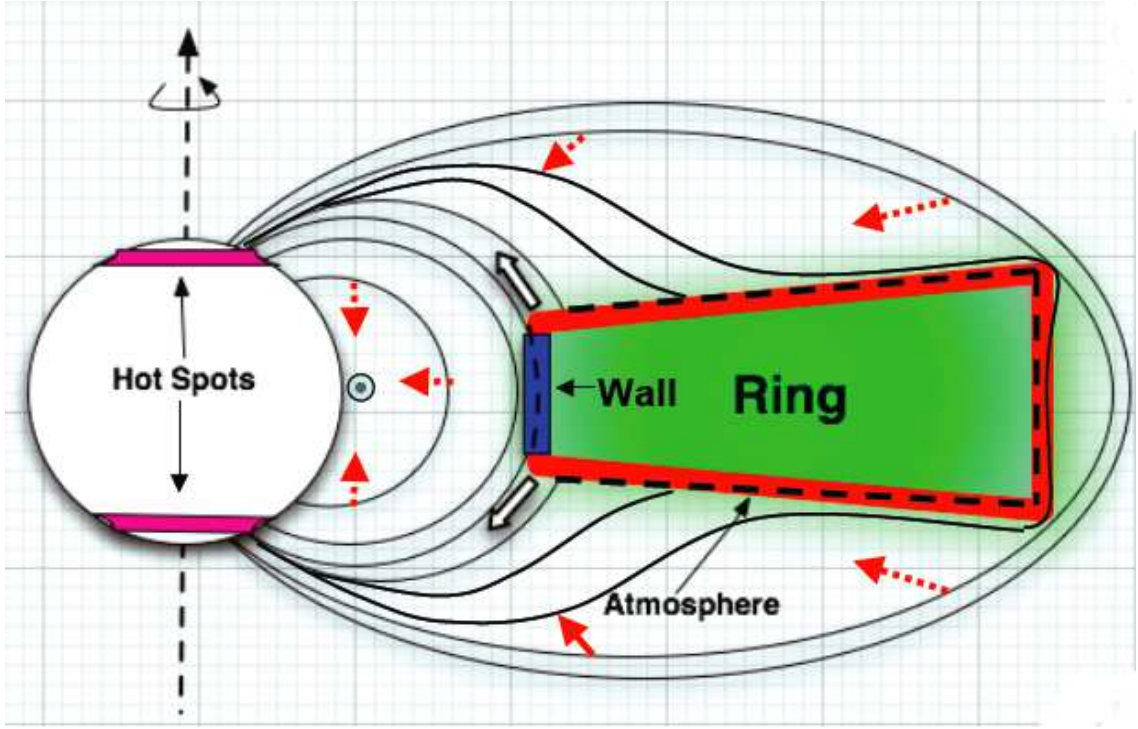}
\end{center}
  \caption{The magnetic field geometry for an  old (i.e. $H_{\rm ring} << \Delta R_{\rm ring}$) QS-ring source; the ring extends from $\sim 15$  to $\sim 100$ km radially. The  aligned dipole, shrinks inwards (dotted red arrows) as
the magnetic field dissipates at the star's equator,  and is also stretched by the outward viscous spreading
 of the ring. 
 Unlike the highly conducting degenerate part of the ring, the non-degenerate atmosphere is quickly penetrated (on timescales of $\sim$ years)  by the  magnetic field (dashed
line) and forced to co-rotate. Accretion (leakage) of the ring's atmosphere ensues at $R_{\rm in}$ as depicted
  by the white arrows. The emission from the corresponding hot spots (grey bands) yields 
    the quiescent phase emission in our model. The magnetic field penetration of the inner wall occurs every 
   100 to 1000 years (depending on the system's age) and  leads to the bursting phase in our model.}
   \label{fig:illustration}
\end{figure*}

       All of  QS-Ring systems should eventually enter an accretion-dominated phase.  
        The onset of accretion  is dictated by an interplay between the
         magnetic field and the ring/atmosphere:   The highly conducting degenerate ring is not penetrated by the magnetic field.    
   In OLNIV  a ``squashed" field was necessary to explain some aspects of
    transient AXPs,  while in OLNII 
  a ``stretched" configuration (shown in Figure \ref{fig:illustration} here) was appropriate. This suggests
 one evolves into the other as a consequence of ring viscous spreading and magnetic
 field decay.  
 
        \subsection{Bursting}

The bursting properties   of SGRs and AXPs  are caused by the
     interaction between the QN debris material and the star's magnetic field.
      This  leads to accretion of debris onto the QS.   
The fundamental difference
 between the co-rotating shell (SGRs) and the keplerian ring (AXPs) is stability.
 The co-rotating shell is a very unstable
  configuration with the shell's boundary hovering around the magnetic neutrality line prompting more
    instabilities (and thus more recurrent bursts). The Keplerian ring  is
   more stable and is only perturbed
     after magnetic field penetration of its inner edge, which triggers 
     the accretion of a small inner part of the ring (i.e. the wall).
      High magnetic field strength is not a 
      requirement in our model: Wall penetration can occur even
       in old sources.  Below,  the
                quiescent and bursting properties of \src~ are explained in the
                 context of an old evolved accreting QS-ring system.
                 
\subsection{Emission components}
                 
  QS-shell
     systems (SGRs) and non-accreting QS-ring systems (transient AXPs) 
      in quiescence  consist of  two emitting components
      (QS vortex annihilation and shell/ring reprocessing the QS flux). Accreting QS-ring systems (AXPs), during quiescence, have 
       three components: The accreting hot spot, the QS vortex annihilation and the illuminated ring.    All  types of systems would burst
      when chunks of the shell/ring are accreted onto the star. 
      Only during the bursting phase would transient AXPs resemble
       AXPs  (OLNIV).

     \section{\src~ as an evolved accreting QS-ring system}
     \label{sec:application}
  
  Here we describe the properties    of an accreting QS-ring system
 in the case where the system has evolved to low-mass, low-density and
  different ring geometry. In this section,  equations from our
  previous studies are embedded in the text or un-numbered; while new equations
   are numbered.
        
        \subsection{Ring properties}
        
         Figure \ref{fig:illustration}  shows the magnetic field configuration and  ring
        geometry for an old accreting
       QS-ring system such as \src.      The ring properties are
       described in detail in \S 2.2 of OLNIV. 
       Hereafter, subscripts ``in" and ``out" refer to the inner and outer
       edge of the ring, respectively.
       The ring vertical height at a radius $R$ is
    $H_{\rm ring} =2.68\ {\rm km}  \rho_{\rm ring, 9}^{1/6} R_{\rm 15}^{3/2}$  where 
     the ring density is in units of $10^9$ g cm$^{-3}$.
   The ring radial extent, which defines the ring's
  outer radius $R_{\rm out}$ in our model,  increases by viscous spreading in time as $R_{\rm out}\simeq \Delta R_{\rm ring} \sim  7.8\ {\rm km}\ T_{\rm keV}^{5/4} t_{\rm yrs}^{1/2}$ with the ring's temperature
   and the system's age  in units of keV and years, respectively. 
   The ring average density can be found from 
          $\rho_{\rm ring} =  m_{\rm ring} /(2\pi R_{\rm out}^2 H_{\rm ring, out})$ 
            which gives   $\rho_{\rm ring} \simeq 8\times 10^{7}\ {\rm gm\ cm}^{-3} m_{\rm ring, -7}^{6/7}/ R_{\rm out, 15}^3$; here $m_{\rm ring, -7}$ is the ring's mass at any given time in units of $10^{-7}M_{\odot}$
 (the ring's initial mass is defined as $m_{\rm ring}^0$).
   The
  characteristics of the ring's atmosphere are given in \S 2.3 in OLNIV.
   The  density in the atmosphere is  $\rho_{\rm atm.} = 460\ {\rm gm}\ T_{\rm keV}^{3/2}$,
    its  height $H_{\rm atm., in}=67.3 \ {\rm cm}\ T_{\rm keV} 
 R_{\rm in, 15}^{3/2}/\mu$ with $R_{\rm in}$ expressed
  in units of 15 km; $\mu$ is the mean molecular weight
   of the ring's atmosphere. This implies a total mass in the atmosphere
  of $m_{\rm atm.}\simeq 2\pi \Delta R_{\rm ring}^2\times H_{\rm atm.}\times \rho_{\rm atm.}
  \sim 10^{17}\ {\rm g} R_{15}^{3/2} T_{\rm keV}^5 t_{\rm yrs}/\mu$.
  The sound speed in the atmosphere is $v_{\rm therm.}\simeq 1.7\times 10^{7}\
  {\rm cm\ s}^{-1}\ T_{\rm keV}^{1/2}/\mu^{1/2}$.

   Inclusion of the $T$-dependence of the mean molecular weight is a new feature
   not considered previosuly.
   Since the atmosphere is rich in iron-group elements, we can write the
    mean molecular weight $\mu=56/(N_{\rm e}+1)$ where
$N_{\rm e}$ is the   mean charge
   of iron  given below in three temperatures regimes as measured in experiments (e.g.
   \cite{arnaud85}; see also \cite{kallman82}  and \cite{makishima86}):  \\
   \begin{eqnarray}
   (N_{\rm e}+1)\sim 9.2 ~ \left(\frac{T}{0.1\ {\rm keV}}\right)^{1/4} \quad &{\rm for}&  T < 0.08\ {\rm keV}\\\nonumber
   (N_{\rm e}+1)\sim 10.3~ \left(\frac{T}{0.1\ {\rm keV}}\right)^{2/3}  \ &{\rm for}&  0.08\ {\rm keV} \le T \le 0.4\ {\rm keV}\\\nonumber
   (N_{\rm e}+1)\sim 27  \quad\qquad\qquad\qquad\quad &{\rm for}& T > 0.4\ {\rm keV}\ .
   \end{eqnarray}

        \subsection{Quiescent Phase: {\it The accretion Band}}
        
        The ring leaks out (i.e. accretes) steadily from its atmosphere at the inner radius (see
        Figure \ref{fig:illustration}), 
$R_{\rm in}$, at a 
 rate  $\dot{m}= \rho_{\rm atm.}  v_{\rm therm.} 2\pi R_{\rm in} (2 H_{\rm atm., in})$ and
  accretes onto the surface of the QS creating a hot spot (HS). 
We get
  \begin{equation}
  \dot{m}\simeq 3.2\times 10^{17}\ {\rm g\ s}^{-1}\ \frac{T_{\rm keV}^3 R_{\rm in, 15}^{5/2}}{\mu_{\rm q, 10}^{3/2}}\ .
  \end{equation}
  
  Irradiation of the ring by the HS luminosity ($L_{\rm HS}=\eta \dot{m} c^2$) leads to a feedback mechanism that  
  allows us to estimate the equilibrium temperature of the ring from
  $\Omega_{\rm ring} L_{\rm HS}= A_{\rm ring}\sigma T_{\rm ring}^4$ where $A_{\rm ring}$ is
    the total area of the ring; the
  irradiation solid angle is 
   $\Omega_{\rm ring}$ and depends on the disk geometry and the
   location of the HS on the surface of the star. We take it as a free parameter  (we expect $\Omega_{\rm ring}< H_{\rm ring}/R\sim 0.1 m_{\rm ring, -7}^{1/7} (R/R_{\rm out})^{1/2}$).
    The ring's equilibrium temperature is then
  \begin{equation}
 \mu_{\rm q, 10}^{3/7}~ T_{\rm ring} \sim 3.4\times 10^{-2} \ {\rm keV}\ \frac{\eta_{0.1}^{2/7}R_{\rm in, 15}^{5/7}  \Omega_{\rm ring}^{2/7}}{t_{\rm Myr}^{2/7}} \\ ,
 \label{eq:Tq}
  \end{equation}
   where $\mu_{\rm q, 10}=\mu_{\rm q, 10}(T_{\rm ring})$ is the quiescent 
    mean molecular weight in units of 10; and time is given in units of million years.  Replacing the equilibrium
    temperature in the accretion equation above gives the equilibrium accretion rate
    \begin{equation}
  \dot{m}_{\rm q} \sim 1.3\times 10^{13}\ {\rm g\ s}^{-1}\ \frac{\eta_{0.1}^{6/7}R_{\rm in, 15}^{65/14}  \Omega_{\rm ring}^{6/7}}{\mu_{q, 10}^{39/14}t_{\rm Myr}^{6/7}} \\ .
  \label{eq:mdotq}
  \end{equation}   
  The consumption of the ring is thus given as
   $dm_{\rm ring}/dt = -  \dot{m}_{\rm q}$  which gives 
     \begin{equation}
     m_{\rm ring} = m_{\rm ring}^0 \left( 1-(\frac{t}{\tau_{\rm ring}})^{1/7} \right)\ .
     \end{equation}
    with the ring's lifetime
    \begin{equation}
    \tau_{\rm ring}\sim  10^5\ {\rm yrs}\ \frac{m_{\rm ring, -6}^0 \mu_{\rm q, 10}^{39/2}}{\eta_{0.1}^6 R_{\rm in, 15}^{65/2}\Omega_{\rm ring}^6}\ .
    \end{equation}
     The ring's mass at birth $m_{\rm ring}^0$ is given in units of $10^{-6}M_{\odot}$.
    The condition
     $\tau_{\rm ring} > t_{\rm age}$ puts a constraint on the ring's mass
     at birth, $m_{\rm ring}^0$ where $t_{\rm age}$ is the system's age.

\begin{figure}
\includegraphics[width=0.5\textwidth]{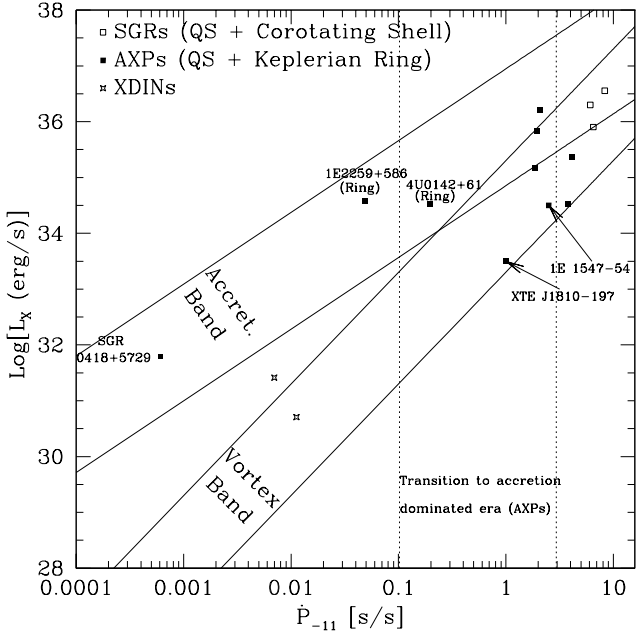}
\caption{The quiescent phase of SGRs, AXPs and XDINs in our model:
 QS-shell systems (i.e. SGRs) evolve along the vortex band, given by 
   magnetic dissipation at the surface of the star following
   vortex expulsion and reconnection. Accreting QS-ring systems
   (i.e. AXPs) evolve along the accretion band given by 
   emission from the hot spot  (see Figure 1).  Transient AXPs (XTE J1810$-$197
   and 1E1547$-$54)
    are   QS-ring systems which have not yet entered the accretion phase and
     evolve along the vortex band.  Old accreting QS-ring systems 
        eventually re-enter the vortex band once they consume their ring.
        Both QS-shell and QS-ring systems  eventually end up at the bottom
          of the vortex band where XDINs are located. }
 \label{fig:bands}
\end{figure}

     The corresponding steady accretion luminosity (defining the
      quiescent phase in our model), $\eta \dot{m}_{\rm q}c^2$, is
     \begin{equation}
  L_{\rm HS, q} \sim 1.2\times 10^{33}\ {\rm erg\ s}^{-1}\ \frac{\eta_{0.1}^{13/7}R_{\rm in, 15}^{65/14}  \Omega_{\rm ring}^{6/7}}{\mu_{q, 10}^{39/14}t_{\rm Myr}^{6/7}}  \\  ,
  \label{eq:lhsq}
  \end{equation}
   
  The above equation can be recast into an $L_{\rm HS}$-$\dot{P}$ form by recalling
   that in our model,  $\dot{P}= \dot{P}_0 (1+t/\tau_0)^{-2/3}\sim \dot{P}_0 (t/\tau_0)^{-2/3}$
    with $\dot{P}_0=P_0/(3\tau_0)$ . We find 
    \begin{equation}
  L_{\rm HS, q} \sim 1.6\times 10^{36}\ {\rm erg\ s}^{-1}\ \frac{\eta_{0.1}^{13/7}R_{\rm in, 15}^{65/14}  \Omega_{\rm ring}^{6/7} }{\mu_{\rm q, 10}^{39/14} (P_{\rm 0, ms}B_{0, 15}^{2})^{3/7}}\times \dot{P}_{-11}^{9/7} \\ .
  \end{equation}
  Shown in Figure \ref{fig:bands} is the accretion Band for two cases of $R_{\rm in}=12$ km (limited
   by the QS radius) and $R_{\rm in}=40$ km (based on fits from previous work).     
   The region between the accretion band and the vortex band 
     is a ``drop" region in our model and defines a regime where
      the ring becomes fully non-degenerate entering a RRAT phase (OLNIII).
      It takes millions of years for the ring to become fully non-degenerate, 
       drastically reducing its accretion rate in the process so that $L_{\rm HS} < L_{\rm v}$.
      RRATs in our model are ring-systems dropping back to the vortex band joining
       the old shell-less and ring-less systems which are XDINs in our model (see \S 8 in OLNI).

         The HS area on the star area can be estimated from eq.(C1) in OLNIV as
          $A_{\rm HS}\simeq 4\pi R_{\rm QS}^2 \left(  \cos \beta_{\rm max}-  \cos \beta_{\rm in} \right)$ 
   with
   \begin{eqnarray}
    \sin \beta_{\rm in} &=& \left( \frac{R_{\rm QS}/R_{\rm in}}{1+ (H_{\rm ring, in}/R_{\rm in})^2}\right)^{1/2}\\\nonumber 
    \sin \beta_{\rm max} &=& \left( \frac{R_{\rm QS}/R_{\rm in}}{1+ (H_{\rm max}/R_{\rm in})^2}\right)^{1/2}\\\nonumber    
    H_{\rm max} &=& H_{\rm ring, in} + H_{\rm atm., in}
     \end{eqnarray}
     Since $H_{\rm atm.} << H_{\rm ring}$,     the above becomes
     \begin{equation}
     A_{\rm HS}\sim  10^8\ {\rm cm}^2\ \frac{R_{\rm QS, 10}^3}{(1-\frac{R_{\rm QS}}{R_{\rm in}})^{1/2}}\ \frac{m_{\rm ring, -7}^{1/7}}{\mu_{\rm q, 10}T_{\rm keV}^{1/4} t_{\rm Myr}^{1/4}} \\ .
     \label{eq:ahs}
     \end{equation}

         \subsection{Bursting phase} 
         
         \subsubsection{Ring penetration and wall accretion}
        
          The inner ring is subject to tidal forces which breaks it into vertical ``walls" of thickness
           $\delta r_{\rm w} = 400\ {\rm cm}\ R_{\rm in, 15}^{3/2}$ (eq. 5 in OLNII).
            The mass of the innermost wall is then $m_{\rm w}= 2\pi R_{\rm in}\times (2 H_{\rm ring, in})\times \delta r_{\rm w}\times \rho_{\rm ring}$.  Including the dependence of $\rho_{\rm ring}$ on
             other parameters we find,
          \begin{equation}
          m_{\rm w} =  7.5\times 10^{-14} M_{\odot} \frac{m_{\rm ring, -7}^{6/7}R_{\rm in, 15}^{7/8}\mu_{q,10}^{15/8}}{\eta_{ 0.1}^{5/4} t_{\rm Myr}^{1/2}} \\ .
          \end{equation}        
          The  bursting phase is initiated when the innermost wall of the Keplerian
         ring is magnetically permeated, and forced to co-rotate
          with the dipolar field, such that it detaches from the ring  and is accreted.
                       The corresponding burst energy (the first burst component), $E_{\rm b}=\eta m_{\rm w}c^2$,  is then
  \begin{equation}
  E_{\rm b}\sim 1.3\times 10^{40} {\rm erg}  \frac{m_{\rm ring, -7}^{6/7}R_{\rm in, 15}^{7/8}\mu_{q, 10}^{15/8}}{\eta_{ 0.1}^{1/4} t_{\rm Myr}^{1/2}} \\ .
  \label{eq:eburst}
  \end{equation}
  In reality, the penetrated wall cracks and breaks into chunks
                   which are then stochastically accreted. 
The wall   is consumed on timescales given by eq.(22) in OLNII which yields 
                  a wall consumption time of a few hours.  This first component of
                   the bursting phase will be defined by sporadic ``spikes".

  \subsubsection{Ring irradiation}          
            
            The second component during the bursting phase consists of the Keplerian ring
             being irradiated by MeV  photons from the accreted wall.  The radiation energy hitting
              the ring is $\Omega_{\rm ring}E_{\rm b}$ which leads to photo-disintegration of
              iron nuclei.    To knock-off a proton from iron takes 10.3 MeV, for an $\alpha$
   it takes 7.6 MeV and all the way down to Aluminum, it takes 26.9 MeV.
    So photo-disintegration of  iron nuclei enriches the 
      atmosphere with light elements with an average $A\sim  28$. 
      This  results in  the decrease of the  molecular weight, $\mu$, of the ring's atmosphere 
               from $\mu_{\rm q}$  to $\mu_{\rm b}$ where subscript ``b" stands for burst.
              The resulting molecular weight is
                  $\mu_{\rm b}\sim 1$ for a proton-rich atmosphere and
                   $\mu_{\rm b}\sim 2.1$ for an atmosphere with any other
                    products of $\gamma$-disintegration except protons;  we adopt
                      an average $\mu_{\rm b}\sim 1.5$.
              
              The maximum amount in mass of iron disintegrated is
              $m_{\rm d, max} = A_{\rm ring} \rho 2 \lambda_{\gamma} = 4\pi R_{\rm out}^2 \rho \lambda_{\gamma}$
               where $A_{\rm ring}$ is the ring's total area and $\lambda_{\gamma}=1/(n \sigma_{\rm Fe})$ is the photon penetration depth
                with the disintegration cross-section of iron nuclei for 10 MeV photons is of the
                order of $10^{-25}$ cm$^2$ (e.g. \cite{rengarajan1973}). 
                The density cancels out and we get
               \begin{eqnarray}
               m_{\rm d, max}&\sim& 2.5\times 10^{27}\ {\rm g}\ \frac{T_{\rm keV}^{5/2}\mu_{\rm q, 10} t_{\rm Myr}}{\sigma_{\rm Fe, -25}}\\\nonumber
               &\sim&  4.1\times 10^{23}\ {\rm g}\ \frac{\eta_{0.1}^{5/7} \Omega_{\rm ring}^{5/7}R_{\rm in,15}^{25/14}t_{\rm Myr}^{2/7}}{\mu_{\rm q, 10}^{1/14}\sigma_{\rm Fe, -25}}\ ,
               \end{eqnarray}
               where we made use of equation (\ref{eq:Tq}).      
               In reality, the mass of disintegrated iron is limited
                by the burst energy: $m_{\rm d}\sim (\Omega_{\rm ring}E_{\rm b}/10\ {\rm MeV})\times 56 m_{\rm H}$ which gives
               \begin{equation}
               m_{\rm d}\sim 1.3\times 10^{23}\ {\rm g}\ \frac{\Omega_{\rm ring} m_{\rm ring, -7}^{6/7}R_{\rm in, 15}^{7/8}\mu_{q, 10}^{15/8}}{\eta_{ 0.1}^{1/4} t_{\rm Myr}^{1/2}} \\ .
               \end{equation}
                      
\begin{figure}
\begin{center}
\includegraphics[width=0.5\textwidth]{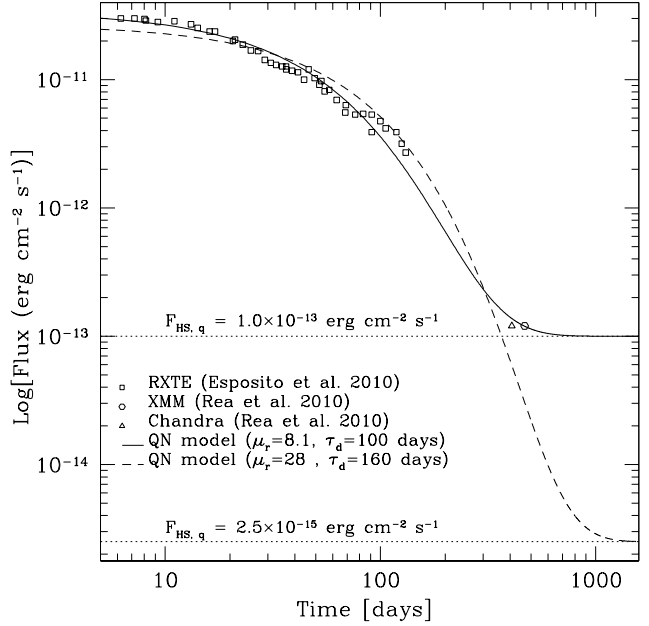}
\end{center}
\caption{Outburst  X-ray flux
 of \src~  as seen by RXTE, XMM and Chandra. Our model's best fit
  is obtained for  $\tau_{\rm d}=100$ days, $\mu_{\rm r}=8.1$ 
 and a  quiescent flux $F_{\rm HS, q}\sim 10^{-13}$ erg cm$^{-2}$ s$^{-1}$
 (i.e.  $L_{\rm HS, q}\sim 4.8\times 10^{31}$ erg s$^{-1}$  for a source at 2kpc).
 Our model gives $L_{\rm HS, q}/ L_{\rm HS, b}=1/\mu_{\rm r}^{39/14}$ at $t_{\rm b}=0$ (start of outburst; eq.(\ref{eq:lhsb})).
  The maximum value of $\mu_{\rm r}=28$ (see text)  implies a minimum 
   quiescent flux of $F_{\rm HS,q}\sim 2.5\times 10^{-15}$  erg cm$^{-2}$ s$^{-1}$
   and requires $\tau_{\rm d}=160$ days for the  best fit (dashed line). }
\label{fig:taud}
\end{figure}

             The atmosphere of the ring evolves back
                to $\mu_{\rm q}$ as (see eq. (27) in OLNII)
                \[
                \frac{1}{\mu(t)} =  \frac{1}{\mu_{\rm q}} + \left(\frac{1}{\mu_{\rm b}} - \frac{1}{\mu_{\rm q}}\right)\exp^{\left(-\frac{t_{\rm b}}{\tau_{\rm d}}\right)}\ ,
                \]
                where the $t_{\rm b}$ means time since the start of burst, during which $t_{\rm Myr}$ is constant ($t_{\rm b} << t_{\rm Myr}$).
                The corresponding HS luminosity\footnote{For younger sources with $\Delta R_{\rm ring}\sim R_{\rm in}$, 
                the ring's solid angle and area have different dependency on ring parameters which
                 leads to $L_{\rm HS, b} \propto (1/\mu)^6$  (eq. 28 in OLNII; see also Figure 2
                 in that paper) instead of the $L_{\rm HS, b} \propto (1/\mu)^{39/14}$
                  for older sources as is the case here.} is 
                 \begin{equation}
                L_{\rm HS, b} =  L_{\rm HS, q}  \left(1+ (\mu_{\rm r}-1) \exp^{\left(-\frac{t_{\rm b}}{\tau_{\rm d}}\right)}\right)^{39/14}\ ,
                \label{eq:lhsb}
                \end{equation}
                where $\mu_{\rm r} =\mu_{\rm q}/\mu_{\rm b}$ and  $\tau_{\rm d}= m_{\rm d}/\dot{m}_{\rm b}$ 
                 is the time it takes the irradiated ring to deplete its light-element-rich atmosphere (through accretion
                 at $R_{\rm in}$).  Here $\dot{m}_{\rm b}$ is given by eq.(\ref{eq:mdotq}) for $\mu = \mu_{\rm b}$.                This yields
                \begin{equation}
                \tau_{\rm d} \sim 189\ {\rm days}\  \frac{\Omega_{\rm ring}^{1/7} m_{\rm ring -7}^{6/7}t_{\rm Myr}^{5/14}\mu_{q, 10}^{15/8}\mu_{\rm b}^{39/14}}{\eta_{0.1}^{31/28}R_{\rm in, 15}^{155/56}} \\ .
                \label{eq:taud}
                \end{equation}
                  Figure \ref{fig:taud} shows our model compared to the observed outburst decay  flux for \src~ (\cite{esposito10}).
                The best fit is obtained for  $\tau_{\rm d}=100$ days, $\mu_{\rm r}=8.1$
                 and a flux during quiescence of $F_{\rm HS, q}\sim 10^{-13}$ erg sm$^{-2}$ s$^{-1}$ (i.e.
                  $L_{\rm HS, q}= 4.8\times 10^{31}$ erg s$^{-1}$ at 2 kpc).

In general it is hard to know the true quiescent
level of many of these objects, since they are observed so rarely when not in outburst.
 Nevertheless,  a minimum value for the flux in quiescence can be obtained
 in our model for the  maximum $\mu_{\rm r}=28$ which corresponds to 
   $\mu_{\rm q}=28$ (i.e. $N_{\rm e}= 1$) and $\mu_{\rm b}=1$ (corresponding to a proton-rich atmosphere). The corresponding 
   best fit (shown as dashed line in Figure \ref{fig:taud}) is obtained for $\tau_{\rm d}= 160$ days
    and $F_{\rm HS, q}=   2.5\times 10^{-15}$ erg sm$^{-2}$ s$^{-1}$. However this $\mu_{\rm r}=28$ 
     fit  misses the most recent measurements.    
     
    Our favored fit above with  $\mu_{\rm r}\sim 8.1$ (i.e. $\mu_{\rm q}\sim 12.2$ 
     for  $\mu_{\rm b}\sim 1.5$) implies an atmosphere temperature
     of $\sim 6.3$ eV (from $\mu\sim 6.1/ T_{\rm 0.1\ keV}^{1/4}$).   Adopting the lower limit of 16 Myr for the
  age of the system ($P/3\dot{P}$), a self-consistent fit 
  with $T_{\rm ring, q}\sim 6.3$ eV (into eq. (\ref{eq:Tq})), and $F_{\rm HS, q}=10^{-13}$ erg m$^{-2}$ s$^{-1}$
   (i.e. $L_{\rm HS, q}\sim 4.8\times 10^{31}$ erg s$^{-1}$ into eq. (\ref{eq:lhsq}))   is obtained for 
    the following current ring properties
  \begin{eqnarray}
   R_{\rm in}&\sim& 36\ {\rm km}\\\nonumber
  R_{\rm out}&\sim&  56\ {\rm km}\\\nonumber
  \Omega_{\rm ring}&\sim& 6.6\times 10^{-3}\\\nonumber
   m_{\rm ring} &\sim& 2.5\times 10^{-10}M_{\odot}\\\nonumber
   \rho_{\rm ring}&\sim& 2\times 10^4\ {\rm gm\ cm}^{-3} \ .
  \end{eqnarray}
   The ring's mass was obtained by equating the observed burst energy 
   of $\sim 6\times 10^{37}$ erg into our eq. (\ref{eq:eburst}). The ring's
   atmosphere has a thickness of $H_{\rm atm.}\sim 0.1$ cm and a mass
    of $m_{\rm atm.}\sim 5\times 10^{12}$ gm.
    The corresponding HS emitting area during  quiescence (eq. \ref{eq:ahs})
   is $A_{\rm HS, q} \sim 8\times 10^7\ {\rm cm}^2$. Also,  eq. (\ref{eq:ahs}) gives
   $A_{\rm HS, q}/A_{\rm HS, b}= (\mu_{\rm b}T_{\rm b, keV}^{1/4})/(\mu_{\rm q}T_{\rm q, keV}^{1/4})$ 
    which implies a HS emitting area during burst of $A_{\rm HS, b} \sim 2\times 10^8\ {\rm cm}^2$. Finally, using the values above in eq. (\ref{eq:taud}) we get $\tau_{\rm d}\sim 0.4\ {\rm days}\
   \mu_{\rm b}^{39/14}$.   Adopting
     an average $\mu$ during accretion of $\mu_{\rm b, av.}\sim (\mu_{\rm q}+\mu_{\rm b})/2\sim 6.8$ gives $\tau_{\rm d}\sim 91$
     days
    which is close to the fit best fit value of 100 days.

\section{Predictions and conclusion}

\begin{enumerate}[1.]
  
  \item  \src's   current ring's mass is $\sim 2.5\times 10^{-10}M_{\odot}$   extending from
   $R_{\rm in}\sim 36$ km to $R_{\rm out}\sim 56$ km (see \ref{fig:illustration}).
    The ring's mass at birth is $m_{\rm ring}^0 > 9\times 10^{-7}M_{\odot}$ found from
     the $\tau_{\rm ring} > t_{\rm age}$ constraint.
    Given their  degenerate nature the rings would have
   a very weak optical signature (unlike what would be expected
    from non-degenerate fall-back disks around neutron stars). 
    
    \item     \src's fit  parameters give $m_{\rm w}\sim 3\times 10^{-16}M_{\odot}$. 
The wall   is consumed on timescales given by eq.(22) in OLNII which yields 
                  a wall consumption time of $< 1$ hour for \src.
                  
   \item Our model predicts a glitch, during the bursting phase (see eq.(29) in OLNII), of 
   \begin{equation}
  \frac{\Delta P}{P}\sim -1.3\times 10^{-10}
  \frac{P_{10}R_{\rm in, 15}^{23/8}m_{\rm ring, -7}^{6/7}\mu_{\rm q, 3.3}^{15/8}}{I_{45}\eta_{0.1}^{5/4}t_{\rm Myr}^{1/2}}\  ,
   \end{equation}
   where $I_{45}$ is the star's moment of inertia in units of $10^{45}$ gm cm$^2$. 
   Using the best fit  parameters for \src~
    gives  $\Delta P/P\sim -2\times 10^{-11}$.

  \item Fundamentally,  the two-components model (QS and ring)  provides
  a natural explanation for multiple emission components: 
    vortex annihilation, the HS on the QS and the illuminated ring. 
    However, as the system
     ages (i.e. at smaller $\dot{P}$),  the ring becomes thinner in height and more extended in radial
      width and area.  The feedback effect is reduced and so  the  ring's temperature decreases
       which makes the ring's contribution during quiescence harder to detect for old sources. 
        In the case of \src~, and after full recovery from bursting phase, we predict a 2-component spectrum.
   The HS at $\sim 0.67$ keV and the ring at $T_{\rm ring}\sim 6.3$ eV.
   A third, much weaker, component would be related to the emission
    from the magnetic field reconnection at the surface of the QS following vortex 
    expulsion and annihilation. Assuming a BB emission, the temperature of the QS would
     be $T_{\rm QS}\sim 17.3\ {\rm eV}~ \dot{P}_{-11}^{1/2}/R_{\rm QS, 10}^{1/2}$
      or $< 0.4$ eV in the case of \src.
      
         \item During the bursting phase the ring's atmosphere material (in particular protons and $\alpha$ particles)
  will emit a cyclotron line $2\pi \nu_{\rm p}\sim e B_{\rm in}/(m_{\rm H} c)$
   with $B_{\rm in}= B_{\rm s} (R_{\rm QS}/R_{\rm in})^3$ and $B_{\rm s}=\sqrt{3\kappa P \dot{P}}
   \sim 5.2\times 10^{14} P_{10}^{1/2} \dot{P}_{-11}^{1/2}$. 
   Recalling that a 1 keV line corresponds to $2.42\times 10^{17}$ Hz
   this gives
    $\nu_{\rm p}\sim 1\ {\rm keV} P_{10}^{1/2} \dot{P}_{-11}^{1/2} R_{\rm in, 15}^{-1/2}$.
    For \src~ we get $\nu_{\rm p}\le 0.05$ keV and $\nu_{\alpha}= 0.5\nu_{\rm p}\le 0.025$ keV.

 \item During quiescence, the ring atmosphere (and thus the accreted
  material), is composed mostly of  pure iron group nuclei.
 However, during the bursting phase and after irradiation,  the atmosphere should
  be composed  mostly of protons, of $\alpha$ particles and 
  of ionized nuclei with $A \sim 28$ ($\mu=\mu_{\rm b}\sim 1.5$). It would be interesting
   to look for such signatures in emission during  quiescence and bursting episodes.

\item A key difference between objects showing magnetar-like activity and
    regular pulsars is the lack of persistent radio emission in the former class
     of objects.  In our model the vortices force the interior magnetic
     field to align with the rotation axis, thus inhibiting persistent radio pulsation (\cite{ouyed2004,ouyed2006}).
      However, if  \src~ has entered   the last stages of ring consumption ($t_{\rm age}\sim
       \tau_{\rm ring}$),
       it should eventually show sporadic radio emission as  it makes its way
       back to the vortex band as a RRAT (see OLNIII).  This suggests that the system is 
        in the last $\sim 1$\% of its lifetime and currently descending from the accretion band
        back to the vortex band. 
               
\end{enumerate}

The above listed predictions in general produce weak signals: (i) Predictions 1\&4 
 yield a flux of  $\sim 10^{-19}$ ergs cm$^{-2}$ s$^{-1}$  $\AA^{-1}$ which corresponds to a 26-27 magnitude in the V-band at 2kpc. This is  in principle detectable from eight meter class telescopes.
 The signal from vortex annihilation is much weaker; (ii) for  prediction 2, the
next burst is not expected until of order of 2000 years from now; (iii) for 3
 the glitch  is small compared to ordinary pulsar glitches ($10^{-9}$ to $10^{-7}$)
 requiring timing precision higher than possible for a transient source; (iv) for 
5 the frequency is below the Lyman limit and the line would be absorbed by 
the hydrogen in the ISM; (v) for 6, the flux during outburst
is high enough to check existing X-ray spectra for intermediate elements lines;
for 7, unfortunately the timescale to evolve into a RRAT is of the order of $10^4$-$10^5$ years.
 In summary, predictions 4 and 6 are the  most testable.

\section*{Acknowledgments}

This research is supported by grants from the Natural Science and Engineering
 Research Council of Canada (NSERC).

\bsp

\label{lastpage}

\end{document}